\documentclass[cameraready]{Interspeech}

\title{Parallel Time--Band Mixing with Learned Observation-Adding for Robust ASR Front-Ends}

\author[affiliation={1,3}, correspondingauthor]{Xingyu}{Shen}
\author[affiliation={1}]{Runze}{Wang}
\author[affiliation={1}]{Wei-Ping}{Zhu}
\author[affiliation={2}]{Benoit}{Champagne}

\address{
    $^1$ Department of Electrical and Computer Engineering, Concordia University, Canada \\
    $^2$ Department of Electrical and Computer Engineering, McGill University, Canada \\
    $^3$ Artificial Intelligence Research Institute, Shenzhen University of Advanced Technology, China
}
\email{xingyu.shen@mail.concordia.ca, runze.wang@mail.concordia.ca, weiping@ece.concordia.ca, benoit.champagne@mcgill.ca}


\usepackage{comment}
\usepackage{booktabs}
\usepackage{multirow}
\usepackage{graphicx}
\usepackage{amsmath}
\usepackage{amssymb}
\usepackage{float}
\usepackage{placeins}
\usepackage{amsmath}
\let\TIPAbar\|   
\let\|\Vert      

\begin{document}

\maketitle



\begin{abstract}
Speech enhancement is often used as a front-end for robust ASR, yet recurrent temporal and cross-band modules introduce sequential dependencies that reduce parallel efficiency. In this paper, we present a sequence-parallel band-split enhancement front-end built on a Parallel Time--Band Mixer (PTBM) block that eliminates within-block recurrent unrolling. PTBM integrates intra-band temporal mixing and per-frame cross-band attention within a unified parallel architecture, enabling efficient contextual modeling across both time and frequency dimensions. The system retains the mask-plus-residual reconstruction interface and introduces learned Observation-Adding (LOA) to suppress ASR-sensitive artifacts without development-set tuning. Experiments on DNS Challenge and CHiME-4 with frozen Whisper back-ends show that the proposed front-end consistently reduces word error rate relative to recurrent band-split baselines while requiring only 0.96\,M parameters and 0.58\,GMAC/s for the front-end network.
\end{abstract}

\section{Introduction}
Robust automatic speech recognition (ASR) remains challenging under real-world noise and reverberation. Deep learning-based enhancement and separation have therefore become a standard component in robust speech pipelines, especially when the recognizer is fixed or difficult to retrain \cite{wang18_taslp_overview}. A common practice is to place a speech enhancement (SE) model as a front-end to suppress interference before recognition, allowing the ASR back-end to remain fixed. In practice, front-end enhancement is constrained by two factors: (1) enhancement artifacts can degrade recognition performance even when speech quality appears improved; (2) the front-end must remain computationally lightweight to enable efficient inference and deployment.

Several recent works have revisited how to make SE beneficial to ASR rather than only improving perceptual quality. Yang \textit{et al.} proposed an attentive recurrent network (ARN) operating in the time domain, while Dissen \textit{et al.} trained a small front-end adapter with ASR gradients to better match a frozen recognizer \cite{yang23_interspeech,dissen24_interspeech}. Hu \textit{et al.} addressed over-suppression via dual-path style learning, and Luo \textit{et al.} proposed pseudo-supervision to adapt SE models to real far-field recordings for downstream speech recognition \cite{hu23_interspeech,luo25c_interspeech}. More broadly, sequence-parallel temporal convolution and attention models have shown that long-context modeling can be achieved without recurrent unrolling, motivating mixer-style designs for deployable speech front-ends \cite{luo19_convtasnet,vaswani17_attention}.

These efforts have revealed that ASR is sensitive to the type of enhancement error. Iwamoto \textit{et al.} decomposed enhancement errors and showed that artifact components can dominate ASR degradation; they also demonstrated that Observation-Adding (OA), which blends the observed and enhanced signals, can mitigate artifacts and improve recognition performance \cite{iwamoto22_interspeech}. Guan \textit{et al.} proposed loss designs that explicitly reduce speech distortion and artifacts in enhancement outputs, while de Oliveira \textit{et al.} cautioned that over-optimizing a single enhancement metric can be misleading for recognition performance, highlighting the need for design choices aligned with downstream ASR objectives \cite{guan24_interspeech,deoliveira24_interspeech}.

 Band-split enhancement architectures provide a strong low-distortion backbone for ASR front-ends. Yu \textit{et al.} introduced Band-split RNN (BSRNN), which alternates temporal modeling and cross-band modeling to improve full-band reconstruction quality \cite{yu23b_interspeech}. To reduce the overhead of such front-ends, Zhao \textit{et al.} proposed frame resampling and progressive sub-band pruning on top of BSRNN to lower computation while maintaining robust ASR performance \cite{zhao25h_interspeech}. Despite these advances, BSRNN-style front-ends still rely on recurrent computation for both temporal and cross-band modeling, which introduces sequential dependencies and reduces parallel efficiency. Meanwhile, recent work on multichannel enhancement suggests that parallelizable sequence modeling with cross-domain interaction is a promising direction for efficient utterance-level systems \cite{shen25_taslpro_dualpath}.

Motivated by the above considerations, we propose a sequence-parallel band-split SE front-end for robust ASR. Our contributions are threefold:
\begin{itemize}
\item We develop the Parallel Time–Band Mixer (PTBM), which combines intra-band temporal convolutional mixing and per-frame cross-band self-attention in a unified parallel architecture, eliminating within-block sequential dependencies and enabling efficient sequence-parallel processing.
\item We introduce learned Observation-Adding (LOA), a lightweight adaptive blending mechanism that automatically predicts OA coefficients, thereby eliminating the need for development-set tuning.
\item On DNS Challenge and CHiME-4 with frozen Whisper back-ends, the proposed front-end consistently reduces WER relative to recurrent band-split baselines while requiring only 0.96\,M parameters and 0.58\,GMAC/s.
\end{itemize}

\section{Proposed Method}
\label{sec:method}

\begin{figure*}[t]
  \centering
  \includegraphics[width=0.9\linewidth]{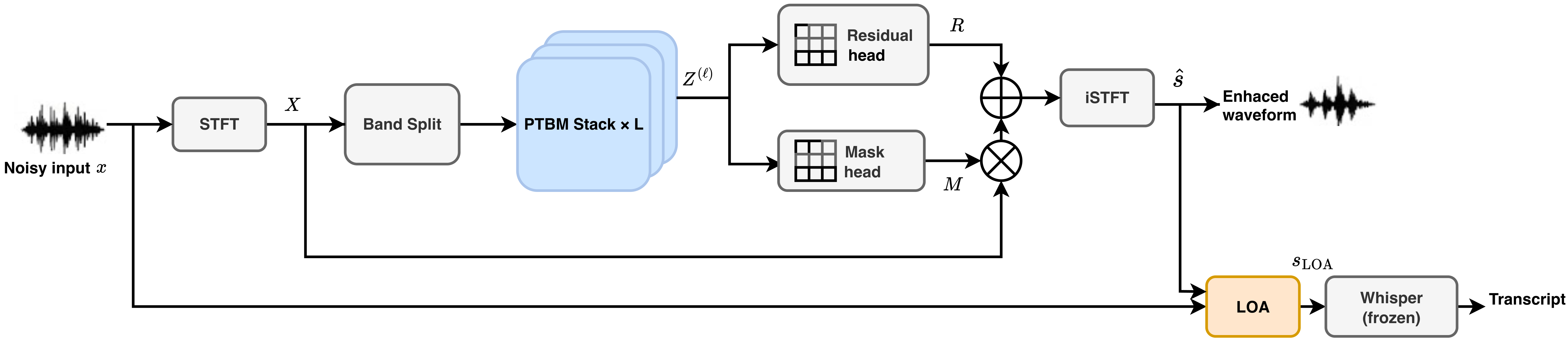}
  \caption{End-to-end pipeline of the proposed speech-enhancement front-end for robust ASR. The system performs band-split time--band modeling with stacked mixer blocks, reconstructs an enhanced waveform using a mask-plus-residual interface, and applies LOA before a frozen Whisper recognizer.}
  \label{fig:overview}
\end{figure*}

\subsection{Overall Architecture}
\label{sec:overall}
Figure~\ref{fig:overview} shows the end-to-end pipeline.
Given a noisy single-channel waveform $x$, we compute its complex STFT $X$ and partition the frequency bins of $X$ into $K$ non-overlapping sub-bands using a predefined (non-uniform) V4 partition \cite{zhao25h_interspeech}.
For each sub-band, we form a real-valued feature by concatenating the real and imaginary parts and applying a band-wise linear embedding, yielding a band-split representation $Z^{(0)}$.
We process $Z^{(0)}$ using a stack of $L$ Parallel Time--Band Mixer (PTBM) blocks to obtain $Z^{(L)}$.
Each PTBM block uses gated dilated depthwise temporal convolutions for intra-band temporal mixing (TCM) and per-frame self-attention over the $K$ sub-bands for cross-band interaction (CBA), avoiding within-block recurrent unrolling.
Two prediction heads map $Z^{(L)}$ to a complex mask $M$ and a complex residual $R$, forming the mask-plus-residual reconstruction interface.
Applying inverse STFT yields an enhanced waveform $\hat{s}$.
Finally, the LOA module predicts a blending weight $\omega\in[0,1]$ from signal statistics and produces $s_{\mathrm{LOA}}$, a weighted sum of the input $x$ and the enhanced estimate $\hat{s}$, which is then fed to a fixed ASR back-end with the recognizer kept frozen.

\subsection{Band Split and Reconstruction}
\label{sec:bandsplit}
Given the complex STFT $X \in \mathbb{C}^{T \times F}$, where $T$ and $F$ denote the numbers of time frames and frequency bins, respectively, we partition the frequency axis into $K$ non-overlapping sub-bands with predefined (non-uniform) sizes $\{F_k\}_{k=1}^{K}$ satisfying $\sum_{k=1}^{K}F_k=F$.
We follow the V4 configuration in~\cite{yu23b_interspeech,zhao25h_interspeech}.
Let $X_k \in \mathbb{C}^{T \times F_k}$ denote the sub-band spectrogram extracted from $X$ by selecting the bins assigned to sub-band $k$.
For each sub-band and time frame, we concatenate the real and imaginary parts of the STFT values across the $F_k$ bins to form a real-valued feature and apply a band-specific linear layer to obtain a $C$-channel embedding.
Stacking the $K$ embeddings over time yields $Z^{(0)} \in \mathbb{R}^{B \times K \times T \times C}$, where $B$ is the batch size.

We adopt the mask-plus-residual reconstruction interface used in BSRNN-style front-ends~\cite{zhao25h_interspeech}.
After processing $Z^{(0)}$ with $L$ PTBM blocks, each sub-band feature is passed through band-specific fully connected layers to predict the real and imaginary parts of the complex mask and residual for the bins in that sub-band.
Concatenating the predictions from all sub-bands yields $M,R \in \mathbb{C}^{T \times F}$ with the same shape as $X$.
The enhanced spectrum is reconstructed by
\begin{equation}
\hat{S} = M \odot X + R,
\label{eq:mr_recon}
\end{equation}
where $\odot$ denotes element-wise multiplication.

\subsection{Parallel Time--Band Mixer Block (PTBM)}
\label{sec:ptbm}
\begin{figure}[ht]
  \centering
  \includegraphics[width=0.6\linewidth]{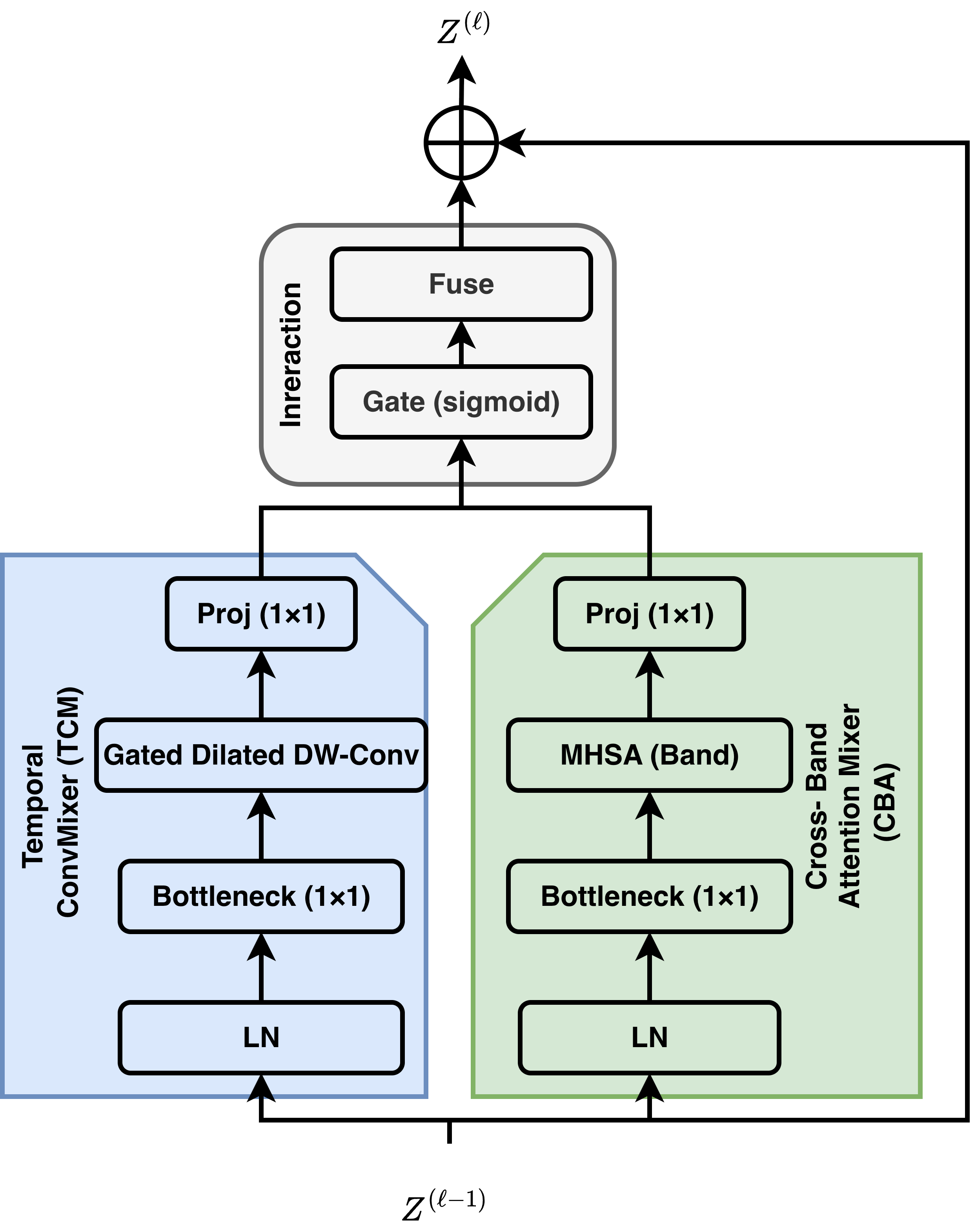}
  \caption{Structure of the PTBM block. The TCM branch and the CBA branch run in parallel, followed by the interaction module and a residual connection.}
  \label{fig:ptbm}
\end{figure}

The core component of the proposed front-end is the PTBM block, whose internal operation is detailed in Figure~\ref{fig:ptbm}.
Each block takes $Z^{(\ell-1)} \in \mathbb{R}^{B \times K \times T \times C}$ as input and outputs $Z^{(\ell)}$ with the same shape, where $\ell\in \{1,2,\cdots,L\}$
indexes the block.
PTBM contains two branches that operate in parallel on $Z^{(\ell-1)}$: a Temporal ConvMixer (TCM) branch for intra-band temporal mixing and a Cross-Band Attention Mixer (CBA) branch for per-frame cross-band interaction.
An interaction module combines the two branch outputs, and a block-level residual connection produces $Z^{(\ell)}$.

\subsubsection{Temporal ConvMixer (TCM)}
\label{sec:tcm}
The TCM branch performs temporal-context mixing within each sub-band.
For each sub-band independently, we apply layer normalization (LN) followed by a point-wise bottleneck projection from $C$ to $C_b$, where $C_b$ is the bottleneck channel dimension.
We then apply gated dilated depthwise 1D convolutions along the frame (time) index $t$ (i.e., the $T$ dimension in $Z^{(\ell-1)} \in \mathbb{R}^{B \times K \times T \times C}$), independently for each example in the batch and each sub-band, to capture temporal context with low computational overhead.
A point-wise projection maps the features back to $C$, yielding the temporal-branch output $Z_{\text{time}}$.

\subsubsection{Cross-Band Attention Mixer (CBA)}
\label{sec:cba}
The CBA branch performs cross-band interaction at each time frame via multi-head self-attention~\cite{vaswani17_attention}.
For each frame $t$, the $K$ sub-band vectors are treated as a token sequence.
After layer normalization (LN), we apply a point-wise bottleneck projection from $C$ to $C_b$, compute multi-head self-attention across the $K$ tokens, and apply a point-wise projection back to $C$ to obtain the cross-band output $Z_{\text{band}}$.
Because attention is applied independently for each frame, this branch is applied in parallel over time frames.

\subsubsection{Interaction}
\label{sec:interaction}
The interaction module in Figure~\ref{fig:ptbm} fuses temporal and cross-band cues using element-wise gating followed by fusion.
Let $Z_{\text{time}}$ and $Z_{\text{band}}$ denote the outputs of the temporal and cross-band branches, respectively.
We compute the gate and the fused representation by
\begin{equation}
\label{eq:fusion}
\begin{aligned}
G &= \sigma\bigl(\operatorname{Linear}([Z_{\text{time}};\, Z_{\text{band}}])\bigr), \\
Z_{\text{fuse}} &= G \odot Z_{\text{time}} + (1 - G) \odot Z_{\text{band}}, \\
Z^{(\ell)} &= Z^{(\ell-1)} + Z_{\text{fuse}} .
\end{aligned}
\end{equation}
where $[\cdot;\cdot]$ concatenates features along the channel dimension, $\operatorname{Linear}(\cdot)$ is a point-wise projection shared across time frames and sub-bands, and $\sigma(\cdot)$ is a sigmoid.

\subsection{Learned Observation-Adding (LOA)}
\label{sec:loa}

LOA is implemented as a two-layer multilayer perceptron (MLP) with a ReLU hidden activation and a sigmoid output, taking summary statistics computed from the observed waveform $x$ and the enhanced waveform $\hat{s}$ as input.
During training, these statistics are computed on the same fixed-length training segment (random crop) used as the SE input; during inference, they are computed on the full-length test recording.

The input statistics consist of (i) a log-energy ratio $\rho_E=\log\frac{\|x\|_2^2+\epsilon}{\|\hat{s}\|_2^2+\epsilon}$ and (ii) the mean and variance of a log-magnitude spectral difference; here $\epsilon$ is a small constant for numerical stability (e.g., $10^{-8}$).
Using the same analysis STFT as in Section~\ref{sec:bandsplit}, we compute the log-magnitude spectral difference between the observed spectrum $X$ and the enhanced spectrum $\hat{S}$ from Eq.~(\ref{eq:mr_recon}) as $\Delta_{t,f}=\log(|X_{t,f}|+\epsilon)-\log(|\hat{S}_{t,f}|+\epsilon)$.
We then compute $\mu_\Delta=\mathrm{mean}_{t,f}(\Delta_{t,f})$ and $\sigma_\Delta^2=\mathrm{var}_{t,f}(\Delta_{t,f})$ over all time--frequency bins of the current input (training segment or test recording).
We concatenate $(\rho_E,\mu_\Delta,\sigma_\Delta^2)$ as the MLP input and predict the blending weight $\omega\in [0,1]$ via a sigmoid output.
These statistics capture both energy mismatch and spectral mismatch and prevent degenerate solutions that only rescale the waveform amplitude.
LOA uses the resulting scalar weight to synthesize the enhanced speech fed to the ASR:
\begin{equation}
s_{\mathrm{LOA}} = \omega \, x + (1-\omega)\, \hat{s}.
\label{eq:loa}
\end{equation}

LOA is trained in a second stage using oracle blending weights.
We first train the SE network with the signal-level objective in Section~\ref{sec:objective}.
We then freeze the SE network and, for each training segment, compute a signal-level oracle weight $\omega^{\ast}\in[0,1]$ by a 1D grid search over the interval $[0,1]$ with a resolution of 0.01. The oracle weight is selected as the value that minimizes the same signal-level loss between the blended waveform $s_{\mathrm{blend}}(\omega)=\omega x + (1-\omega)\hat{s}$ and the clean reference waveform $s$.
The LOA MLP is trained to regress $\omega^{\ast}$ from the segment-level statistics using the $\ell_2$ loss, which keeps LOA lightweight and recognizer-independent.
At inference time, LOA predicts a single blending weight $\omega$ for each test recording, with the input statistics computed over all STFT frames of that recording; thus, the present LOA formulation is utterance-level rather than fully streaming.

\subsection{Training Objective}
\label{sec:objective}
We minimize a fixed weighted sum of the multi-resolution STFT magnitude loss and the SI-SNR loss between the enhanced waveform $\hat{s}$ and the clean reference $s$, following common practice in speech enhancement and separation~\cite{luo19_convtasnet,yamamoto2020parallel}:
\begin{equation}
\mathcal{L} = \mathcal{L}_{\text{MR-STFT}}(\hat{s}, s) + \lambda \, \mathcal{L}_{\text{SI-SNR}}(\hat{s}, s),
\label{eq:loss_total}
\end{equation}
where $\lambda$ is a fixed weight.
We use the standard MR-STFT magnitude loss definition in~\cite{yamamoto2020parallel} and the negative SI-SNR objective as in~\cite{luo19_convtasnet}.
This objective is fully compatible with the mask-plus-residual reconstruction interface in Eq.~(\ref{eq:mr_recon}).

\section{Experimental Setup}
\label{sec:exp}

\subsection{Datasets}
\label{sec:datasets}
We evaluate the proposed SE front-end on two English benchmarks sampled at 16~kHz: DNS Challenge \cite{dns_challenge} and CHiME-4 \cite{chime4}. Following the evaluation protocol in prior band-split front-end studies \cite{zhao25h_interspeech}, we train an SE front-end separately for each benchmark and report results on the corresponding official test sets.

For CHiME-4, we report results on the official single-channel real-recorded development and evaluation sets, \textit{dt05\_real} and \textit{et05\_real} (1ch track) \cite{chime4}. Following the online simulation protocol used in \cite{zhao25h_interspeech}, we generate training mixtures on-the-fly by mixing utterances from \textit{tr05\_simu\_1ch} with noise segments sampled from the DNS noise corpus \cite{dns_challenge}. For each utterance, a noise segment is randomly cropped to match the utterance duration and mixed at an SNR uniformly sampled from $[-5, 20]$~dB.

For DNS Challenge, we follow the ESPnet-SE recipe \cite{espnet} to construct a 100-hour single-channel noisy-speech training set from the official DNS resources \cite{dns_challenge}. Model selection is performed on the official DNS validation set, and we report results on the official DNS synthetic test sets under two conditions: without reverberation and with reverberation \cite{zhao25h_interspeech}.

\begin{table*}[t]
\centering
\caption{Main ASR results and front-end complexity on DNS Challenge and CHiME-4. WER (\%) is reported using frozen Whisper back-ends (Tiny/Medium/Large). Params (M) and MACs (GMAC/s) are reported for the SE front-end only.}
\label{tab:combined_results}
\scriptsize
\setlength{\tabcolsep}{5pt}
\begin{tabular}{l c c ccc ccc ccc ccc}
\toprule
\multirow{3}{*}{\textbf{Method}} & \multirow{3}{*}{\textbf{Params}} & \multirow{3}{*}{\textbf{MACs}} &
\multicolumn{6}{c}{\textbf{DNS Challenge}} & \multicolumn{6}{c}{\textbf{CHiME-4}} \\
\cmidrule(lr){4-9} \cmidrule(lr){10-15}
& & &
\multicolumn{3}{c}{Without Reverb} & \multicolumn{3}{c}{With Reverb} &
\multicolumn{3}{c}{Dev (dt05\_real)} & \multicolumn{3}{c}{Eval (et05\_real)} \\
\cmidrule(lr){4-6} \cmidrule(lr){7-9} \cmidrule(lr){10-12} \cmidrule(lr){13-15}
& \textbf{(M)} & \textbf{(GMAC/s)} &
Tiny & Medium & Large & Tiny & Medium & Large & Tiny & Medium & Large & Tiny & Medium & Large \\
\midrule
Noisy & -- & -- &
15.02 & 8.85 & 4.80 &
39.51 & 21.13 & 10.94 &
21.76 & 5.55 & 4.41 &
35.03 & 8.63 & 6.69 \\
DPARN \cite{hu22_lwfbse,zhao25h_interspeech} & 0.89 & 1.22 &
-- & -- & -- &
-- & -- & -- &
21.13 & 5.52 & 4.67 &
32.97 & 8.84 & 7.56 \\
BSRNN \cite{yu23b_interspeech} & 2.60 & 1.84 &
11.36 & 5.59 & 4.23 &
36.60 & 15.34 & 10.29 &
16.50 & 5.19 & 4.25 &
26.24 & 8.17 & 6.36 \\
Zhao \textit{et al.} \cite{zhao25h_interspeech} & 1.55 & 0.62 &
11.76 & 5.72 & 4.40 &
36.87 & 15.61 & 10.90 &
16.72 & 5.31 & 4.35 &
27.12 & 8.30 & 6.40 \\
\midrule
\textbf{Ours} & 0.96 & 0.58 &
\textbf{11.33} & \textbf{5.58} & \textbf{4.17} &
\textbf{36.35} & \textbf{15.12} & \textbf{10.06} &
\textbf{16.38} & \textbf{5.13} & \textbf{4.18} &
\textbf{26.07} & \textbf{8.11} & \textbf{6.24} \\
\bottomrule
\end{tabular}
\end{table*}

\subsection{Training and Decoding}
\label{sec:traincfg}
All SE front-ends are implemented in ESPnet~\cite{espnet} with the V4 band partition ($K{=}23$)~\cite{yu23b_interspeech,zhao25h_interspeech} and the STFT analysis (32~ms Hann, 16~ms hop, 512-point FFT at 16~kHz).
Unless noted, our model uses $L{=}12$, $C{=}128$, $C_b{=}48$, TCM kernel size 3 with dilations $\{1,2,4,8\}$, CBA with $H{=}4$, and a two-layer LOA MLP (hidden 64, sigmoid output).

For each benchmark, we train a separate SE front-end for 54k steps using Adam~\cite{kingma2015adam} (lr $2\times10^{-4}$, $\beta_1{=}0.9$, $\beta_2{=}0.999$) on 4~s random crops (batch 8) with Eq.~(\ref{eq:loss_total}) ($\lambda{=}1.0$).
MR-STFT uses three resolutions (FFT/hop/win: 1024/120/600, 2048/240/1200, 512/50/240)~\cite{yamamoto2020parallel}.
We select the model with the lowest validation loss on \textit{dt05\_real} (CHiME-4) or the DNS validation set, then freeze the SE network and train LOA for 10k steps with Adam (lr $1\times10^{-4}$) using an $\ell_2$ regression loss to oracle weights from the grid search in Section~\ref{sec:loa}.

We use Whisper~\cite{whisper} as a frozen ASR back-end (Tiny/Medium/Large) and decode without an external language model using the same decoding and text normalization for all methods~\cite{zhao25h_interspeech} (lowercasing, punctuation removal, and whitespace normalization).
Unless noted, ASR input is $s_{\mathrm{LOA}}$; baselines use OA with $\omega$ tuned on the corresponding development set~\cite{iwamoto22_interspeech,zhao25h_interspeech}.
We report Word Error Rate (WER) as the primary ASR metric, and quantify front-end efficiency using Params and MACs (GMAC/s) for the SE network only.

\section{Results and Discussion}
\label{sec:results}

\subsection{Main results on DNS Challenge and CHiME-4}
Table~\ref{tab:combined_results} reports WER and front-end complexity on DNS Challenge and CHiME-4 using frozen Whisper back-ends with three model sizes (Tiny/Medium/Large). Across all back-ends, the proposed front-end achieves lower WER than the noisy input and both band-split baselines (BSRNN \cite{yu23b_interspeech} and the lightweight front-end of Zhao \textit{et al.} \cite{zhao25h_interspeech}). The improvements are consistent on DNS (with/without reverberation) and transfer to real-recorded CHiME-4 sets (dt05\_real and et05\_real), demonstrating robustness beyond the synthetic mixtures used for training.

The table also highlights the accuracy--efficiency trade-off. Compared with BSRNN, our model achieves lower WER with substantially fewer parameters and lower MACs. Compared with Zhao \textit{et al.} \cite{zhao25h_interspeech}, our model improves WER at similar or lower MACs. 

\subsection{Ablation study}
Table~\ref{tab:ablation} reports an ablation study using Whisper Large. Unless otherwise noted, LOA is enabled to isolate architectural changes in the enhancement front-end.

\begin{table}[ht]
\centering
\caption{Ablation study with Whisper Large. We analyze the contribution of different components and the effect of model depth $L$. ``Proposed (Full)'' uses $L{=}12$, TCM, CBA, and LOA.}
\label{tab:ablation}
\scriptsize
\setlength{\tabcolsep}{0.6pt}
\begin{tabular}{l c c c c c}
\toprule
\textbf{Configuration} & \textbf{Params} & \textbf{MACs} & \textbf{DNS} & \textbf{DNS} & \textbf{CHiME-4} \\
 & \textbf{(M)} & \textbf{(GMAC/s)} & \textbf{w/o Rev} & \textbf{w/ Rev} & \textbf{Eval} \\
\midrule
\textbf{Proposed (Full)} & 0.96 & 0.58 & 4.17 & 10.06 & 6.24 \\
\quad w/o CBA & 0.70 & 0.49 & 4.62 & 11.15 & 6.88 \\
\quad w/o TCM & 0.76 & 0.35 & 5.08 & 12.37 & 7.14 \\
\quad w/o OA/LOA (use $\hat{s}$) & 0.96 & 0.58 & 4.45 & 10.65 & 6.60 \\
\quad w/ Fixed OA ($\omega=0.2$ on $x$) & 0.96 & 0.58 & 4.26 & 10.21 & 6.42 \\
\midrule
\multicolumn{6}{l}{\textit{Model Depth Scaling}} \\
\quad $L=6$ & 0.53 & 0.31 & 4.92 & 11.83 & 6.91 \\
\quad $L=9$ & 0.74 & 0.44 & 4.48 & 10.67 & 6.53 \\
\quad $L=15$ & 1.18 & 0.72 & 4.12 & 9.89 & 6.17 \\
\bottomrule
\end{tabular}
\end{table}

Removing cross-band interaction (w/o CBA) consistently degrades WER on DNS and CHiME-4, indicating that explicit per-frame cross-band modeling helps preserve band-to-band structure that is important for recognition.
Removing temporal mixing (w/o TCM) leads to a larger degradation, confirming that strong intra-band temporal context remains crucial for suppressing non-stationary interference.
Together, these results support the PTBM design that combines temporal and cross-band mixing within each block.
Using the enhanced waveform directly as ASR input (w/o OA/LOA; use $\hat{s}$) yields worse WER than applying OA, consistent with prior observations that ASR is sensitive to enhancement artifacts \cite{iwamoto22_interspeech}.
A fixed OA weight improves robustness, while LOA achieves the best performance without development-set coefficient tuning, simplifying deployment across varying conditions.
Increasing the number of PTBM blocks $L$ improves WER with diminishing returns.
Moving from $L{=}6$ to $L{=}12$ yields substantial gains at moderate additional cost, while $L{=}15$ provides only a small further improvement.
We therefore use $L{=}12$ as the default configuration to balance recognition accuracy and front-end complexity.



\subsection{Operator-level analysis}
Table~\ref{tab:operator} compares the temporal and cross-band operators in recurrent band-split blocks and in PTBM under the default configuration in Section~\ref{sec:traincfg}.
TCM and CBA are sequence-parallel and have lower per-invocation cost than gated recurrent unit (GRU)-based temporal/cross-band modules, which aligns with the front-end MAC reduction reported in Table~\ref{tab:combined_results}.

\begin{table}[ht]
\centering
\caption{Operator complexity of temporal and cross-band modules under the default configuration in Section~\ref{sec:traincfg}. Params and MACs are reported per module invocation (temporal modules per sub-band; cross-band modules per time frame).}
\label{tab:operator}
\scriptsize
\setlength{\tabcolsep}{6pt}
\begin{tabular}{l c c c}
\toprule
\textbf{Module} & \textbf{Params} & \textbf{MACs} & \textbf{Seq.-parallel} \\
\midrule
Temporal GRU & 99k & 98k & No \\
Temporal ConvMixer (TCM) & 17k & 17k & Yes \\
\midrule
Cross-band GRU & 99k & 2.26M & No \\
Cross-Band Attn. Mixer (CBA) & 22k & 0.55M & Yes \\
\bottomrule
\end{tabular}
\end{table}

\section{Conclusion}
\label{sec:conclusion}
We presented a band-split sequence-parallel SE front-end built on PTBM blocks that combines temporal convolution mixing and per-frame cross-band attention without recurrent unrolling. 
We also introduced LOA, an artifact-aware blending mechanism that eliminates the need for development-set coefficient tuning.
Across DNS Challenge and CHiME-4 benchmarks with frozen Whisper back-ends, the proposed front-end consistently improves WER over recurrent band-split baselines while reducing model size and MACs. 
These results indicate that parallel time--band mixing is a practical design solution for efficient speech-enhancement front-ends in robust ASR pipelines.

\section{Generative AI Use Disclosure}
Generative AI tools were used only for language editing and stylistic polishing of the manuscript text. They were not used to generate the scientific ideas, methods, experiments, figures, tables, or results. All authors take full responsibility for the content of this paper.

\bibliographystyle{IEEEtran}
\bibliography{mybib}

@inproceedings{zhao25h_interspeech,
  title     = {Lightweight Front-end Enhancement for Robust {ASR} via Frame Resampling and Sub-Band Pruning},
  author    = {Zhao, Siyi and Wang, Wei and Qian, Yanmin},
  booktitle = {Proc. Interspeech},
  year      = {2025},
  month     = aug,
  pages     = {3409--3413},
  doi       = {10.21437/Interspeech.2025-1668}
}

@inproceedings{yu23b_interspeech,
  title     = {High Fidelity Speech Enhancement with Band-split {RNN}},
  author    = {Yu, Jianwei and Chen, Hangting and Luo, Yi and Gu, Rongzhi and Weng, Chao},
  booktitle = {Proc. Interspeech},
  year      = {2023},
  month     = aug,
  pages     = {2483--2487},
  doi       = {10.21437/Interspeech.2023-1433}
}

@inproceedings{iwamoto22_interspeech,
  title     = {How Bad Are Artifacts?: Analyzing the Impact of Speech Enhancement Errors on {ASR}},
  author    = {Iwamoto, Kazuma and Ochiai, Tsubasa and Delcroix, Marc and Ikeshita, Rintaro and Sato, Hiroshi and Araki, Shoko and Katagiri, Shigeru},
  booktitle = {Proc. Interspeech},
  year      = {2022},
  month     = sep,
  pages     = {5418--5422},
  doi       = {10.21437/Interspeech.2022-318}
}

@inproceedings{guan24_interspeech,
  title     = {Reducing Speech Distortion and Artifacts for Speech Enhancement by Loss Function},
  author    = {Guan, Haixin and Dai, Wei and Wang, Guangyong and Tan, Xiaobin and Li, Peng and Liang, Jiaen},
  booktitle = {Proc. Interspeech},
  year      = {2024},
  month     = sep,
  pages     = {1730--1734},
  doi       = {10.21437/Interspeech.2024-620}
}

@inproceedings{deoliveira24_interspeech,
  title = {The {PESQ}etarian: On the Relevance of {Goodhart}'s Law for Speech Enhancement},
  author    = {{de Oliveira}, Danilo and Welker, Simon and Richter, Julius and Gerkmann, Timo},
  booktitle = {Proc. Interspeech},
  year      = {2024},
  month     = sep,
  pages     = {3854--3858},
  doi       = {10.21437/Interspeech.2024-2051}
}

@inproceedings{yang23_interspeech,
  title     = {Time-Domain Speech Enhancement for Robust Automatic Speech Recognition},
  author    = {Yang, Yufeng and Pandey, Ashutosh and Wang, DeLiang},
  booktitle = {Proc. Interspeech},
  year      = {2023},
  month     = aug,
  pages     = {4913--4917},
  doi       = {10.21437/Interspeech.2023-167}
}

@inproceedings{dissen24_interspeech,
  title     = {Enhanced {ASR} Robustness to Packet Loss with a Front-End Adaptation Network},
  author    = {Dissen, Yehoshua and Yonash, Shiry and Cohen, Israel and Keshet, Joseph},
  booktitle = {Proc. Interspeech},
  year      = {2024},
  month     = sep,
  pages     = {5008--5012},
  doi       = {10.21437/Interspeech.2024-306}
}

@inproceedings{hu23_interspeech,
  title     = {Dual-Path Style Learning for End-to-End Noise-Robust Speech Recognition},
  author    = {Hu, Yuchen and Hou, Nana and Chen, Chen and Chng, Eng Siong},
  booktitle = {Proc. Interspeech},
  year      = {2023},
  month     = aug,
  pages     = {2918--2922},
  doi       = {10.21437/Interspeech.2023-101}
}

@inproceedings{luo25c_interspeech,
  title = {{SuPseudo}: A Pseudo-supervised Learning Method for Neural Speech Enhancement in Far-field Speech Recognition},
  author    = {Luo, Longjie and Li, Lin and Hong, Qingyang},
  booktitle = {Proc. Interspeech},
  year      = {2025},
  month     = aug,
  pages     = {3404--3408},
  doi       = {10.21437/Interspeech.2025-1513}
}

@article{shen25_taslpro_dualpath,
  title   = {Dual-Path State-Space Modeling With Cross-Domain Interaction for Multichannel Speech Enhancement},
  author  = {Shen, Xingyu and Wang, Runze and Zhu, Wei-Ping and Champagne, Benoit},
  journal = {{IEEE}/{ACM} Transactions on Audio, Speech, and Language Processing},
  year    = {2025},
  volume  = {33},
  pages   = {4239--4252},
  doi     = {10.1109/TASLPRO.2025.3618543}
}

@article{wang18_taslp_overview,
  title   = {Supervised Speech Separation Based on Deep Learning: An Overview},
  author  = {Wang, DeLiang and Chen, Jitong},
  journal = {{IEEE}/{ACM} Transactions on Audio, Speech, and Language Processing},
  year    = {2018},
  volume  = {26},
  number  = {10},
  month   = oct,
  pages   = {1702--1726},
  doi     = {10.1109/TASLP.2018.2842159}
}

@article{luo19_convtasnet,
  title = {{Conv}-{TasNet}: Surpassing Ideal Time-Frequency Magnitude Masking for Speech Separation},
  author  = {Luo, Yi and Mesgarani, Nima},
  journal = {{IEEE}/{ACM} Transactions on Audio, Speech, and Language Processing},
  year    = {2019},
  volume  = {27},
  number  = {8},
  month   = aug,
  pages   = {1256--1266},
  doi     = {10.1109/TASLP.2019.2915167}
}

@inproceedings{vaswani17_attention,
  title     = {Attention Is All You Need},
  author    = {Vaswani, Ashish and Shazeer, Noam and Parmar, Niki and Uszkoreit, Jakob and Jones, Llion and Gomez, Aidan N. and Kaiser, Lukasz and Polosukhin, Illia},
  booktitle = {Advances in Neural Information Processing Systems},
  year      = {2017},
  volume    = {30},
  month     = dec,
  pages     = {5998--6008}
}

@inproceedings{dns_challenge,
  title     = {The {INTERSPEECH} 2020 Deep Noise Suppression Challenge: Datasets, Subjective Testing Framework, and Challenge Results},
  author    = {Reddy, Chandan K. A. and Gopal, Vishak and Cutler, Ross and Beyrami, Ebrahim and Cheng, Roger and Dubey, Harishchandra and Matusevych, Sergiy and Aichner, Robert and Aazami, Ashkan and Braun, Sebastian and Rana, Puneet and Srinivasan, Sriram and Gehrke, Johannes},
  booktitle = {Proc. Interspeech},
  year      = {2020},
  month     = oct,
  pages     = {2492--2496},
  doi       = {10.21437/Interspeech.2020-3038}
}

@inproceedings{espnet,
  title     = {{ESPnet}: End-to-End Speech Processing Toolkit},
  author    = {Watanabe, Shinji and Hori, Takaaki and Karita, Shigeki and Hayashi, Tomoki and Nishitoba, Jiro and Unno, Yuya and Yalta Soplin, Nelson Enrique and Heymann, Jahn and Wiesner, Matthew and Chen, Nanxin and Renduchintala, Adithya and Ochiai, Tsubasa},
  booktitle = {Proc. Interspeech},
  year      = {2018},
  month     = sep,
  pages     = {2207--2211},
  doi       = {10.21437/Interspeech.2018-1456}
}

@article{chime4,
  title   = {An Analysis of Environment, Microphone and Data Simulation Mismatches in Robust Speech Recognition},
  author  = {Vincent, Emmanuel and Watanabe, Shinji and Nugraha, Aditya Arie and Barker, Jon and Marxer, Ricard},
  journal = {Computer Speech \& Language},
  year    = {2017},
  volume  = {46},
  month   = nov,
  pages   = {535--557},
  doi     = {10.1016/j.csl.2016.11.005}
}

@article{kingma2015adam,
  title         = {Adam: A Method for Stochastic Optimization},
  author        = {Kingma, Diederik P. and Ba, Jimmy},
  journal       = {arXiv preprint arXiv:1412.6980},
  year          = {2014},
  month         = dec,
  url           = {https://arxiv.org/abs/1412.6980},
  archivePrefix = {arXiv},
  eprint        = {1412.6980},
  primaryClass  = {cs.LG}
}

@inproceedings{whisper,
  title     = {Robust Speech Recognition via Large-Scale Weak Supervision},
  author = {Radford, Alec and Kim, Jong Wook and Xu, Tao and Brockman, Greg and McLeavey, Christine and Sutskever, Ilya},
  booktitle = {Proceedings of the 40th International Conference on Machine Learning},
  year      = {2023},
  month     = jul,
  series    = {Proceedings of Machine Learning Research},
  volume    = {202},
  pages     = {28492--28518},
  publisher = {PMLR}
}

@inproceedings{yamamoto2020parallel,
  title     = {Parallel {WaveGAN}: A Fast Waveform Generation Model Based on Generative Adversarial Networks with Multi-Resolution Spectrogram},
  author    = {Yamamoto, Ryuichi and Song, Eunwoo and Kim, Jae-Min},
  booktitle = {Proc. ICASSP},
  year      = {2020},
  month     = may,
  pages     = {6199--6203},
  doi       = {10.1109/ICASSP40776.2020.9053795}
}

@article{hu22_lwfbse,
  title   = {A light-weight full-band speech enhancement model},
  author  = {Hu, Qinwen and Hou, Zhongshu and Le, Xiaohuai and Lu, Jing},
  journal = {arXiv preprint arXiv:2206.14524},
  year    = {2022},
  doi     = {10.48550/arXiv.2206.14524}
}

\end{document}